\documentclass{elsarticle}
\usepackage[utf8]{inputenc}
\usepackage{graphicx}
\usepackage[usenames,dvipsnames]{color}
\usepackage{natbib}
\usepackage{amsmath}
\usepackage{amssymb}
\usepackage{url}

\newcommand{\Aff}{\mathcal{A}}
\newcommand{\mue}{\tilde{\mu}}
\newcommand{\eA}{\tilde{\Aff}}
\newcommand{\st}{\circ}

\newcommand{\jj}{\mathbf{j}}

\begin{document}
\begin{frontmatter}
\title{On the thermodynamic derivation of Nernst relation}

\author[vscht]{Diego del Olmo\corref{cor1}}
\ead{delolmod at vscht.cz}

\author[vscht,mff]{Michal Pavelka}

\author[vscht]{Juraj Kosek}

\address[vscht]{Department of Chemical Engineering, University of Chemistry and Technology Prague, Technická 5, 16628 Prague 6, Czech Republic}
\address[mff]{Mathematical Institute, Faculty of Mathematics and Physics, Charles University, Sokolovská 83, 186 75 Prague, Czech Republic}
\cortext[cor1]{Corresponding author}

\journal{Journal}

\begin{abstract}
What is the maximum voltage of a cell with a given electrochemical reaction? The answer to this question has been given more than a century ago by Walther Nernst and bears his name. Unfortunately, the assumptions behind the answer have been forgotten by many authors, which leads to wrong forms of the Nernst relation. Such mistakes can be overcome by applying a correct thermodynamic derivation independently of the form in which the reaction is written. The correct form of Nernst relation is important for instance in modelling of vanadium redox flow batteries or zinc-air batteries. In particular, the presence of corrosion can impact the OCV in the case of zinc-air batteries.
\end{abstract}
\end{frontmatter}

\textit{"Thermodynamics is a funny subject. The first time you go through it, you don't understand it at all. The second time you go through it, you think you understand it, except for one or two points. The third time you go through it, you know you don't understand it, but by that time you are so used to the subject, it doesn't bother you anymore." (Arnold Sommerfeld)}

\section{Introduction}

The unpredictable nature of renewable energy sources has attracted a lot of research in supporting technologies for energy storage such as batteries. Consequently, numerous journal articles focused on the modelling of battery systems can be found in the literature, providing an invaluable source of information to new researchers. Such a vast amount of sources may lead to the application of widely used concepts without reminding their general derivation. Nonetheless, some of these concepts may rely on underlying assumptions which can be overseen if not treated carefully.

One of such cases is the open-circuit potential (OCV), also referred as electromotive force (emf), which can be used as a state-of-charge estimator \cite{Skyllas-Kazacos2011} or to identify undesired processes taking place in the system \cite{Zhang2006b}. The OCV is calculated from the so-called Nernst equation. As Nernst acknowledges in his book \cite{Nernst}, it was Helmholtz who first derived a formula relating voltage and concentrations of species in the electrolytes. However, Nernst formulated the relation in a general thermodynamic way, and the Nernst equation has become a cornerstone of electrochemistry. In his derivation, 
the Gibbs energy of the reaction will be equal to the maximal electrical work that can be obtained, c.f. \cite{landau5,PKG}, which determines the OCV. In general, it is enough to just take into account the reaction coefficient of both half-cell reactions (associated to the liquid-solid interphases at both electrodes) to reach correct OCV equation. But, for more complex systems, like those using ion-exchange membranes, a more careful derivation considering all interphases present in the system is required.

This work comprehensively reviews the derivation of the Nernst equation from basic thermodynamics. Furthermore, we show how the use of simplified OCV formulas can lead to incorrect OCV predictions for complex systems which can be avoided using the thermodynamic derivation. Finally, special cases in which multiple reactions can take place at one electrode shifting the potential (e.g. corrosion processes) are also discussed. Novelty of this work does not, of course, lie in the Nernst relation itself, but in recalling the principles behind it in nowadays thermodynamic terminology and in careful application of the principles in particular systems (zinc-air, all-vanadium, hydrogen-bromine).


\section{Thermodynamics origin of open-circuit voltage}
Let us first recall a few fundamental concepts of equilibrium, non-equilibrium thermodynamics and physical chemistry, namely the electric potential, electrochemical potential, electrochemical affinity and open-circuit voltage.

\subsection{Electric potential}\label{sec.elpot}
The electrochemical potential of species $\alpha$ in a solution can be generally expressed as the sum of the chemical potential $\mu_\alpha$ and the molar electrostatic potential energy \cite{levine2008} such that
%
\begin{equation}\label{eq.mue}
\mue_\alpha = \mu_\alpha + z_\alpha F \varphi = \mu^\st_\alpha + RT\ln\underbrace{\left(\gamma_\alpha \frac{b_\alpha}{b^\st}\right)}_{a_\alpha} + z_\alpha F \varphi
\end{equation}

where $\mue^\st_\alpha$ is the standard chemical potential of the species $\alpha$ (here in aqueous solution), $b_\alpha$ is its molality ($b^\st$ the standard molality), $\gamma_\alpha$ is its activity coefficient
, $a_\alpha$ is the activity, $z_\alpha$ is the charge number and $\varphi$ is the electrostatic Maxwell potential, for which the Poisson equation holds in electrostatics, see e.g. \cite{PKG,Kjelstrup}. 


The potential measured by a potentiometer is proportional to the difference in electrochemical potential of electrons between the terminals (measuring electrodes). The electrochemical potential of electrons is defined as

\begin{equation}\label{eq.Phi}
\mue_{e^-} = -F\Phi,
\end{equation}
which should be regarded also as the definition of electric potential $\Phi$, see \cite{Kjelstrup,pavelka-jps,Pavelka-AE} for more details. 

Note that the potentials $\varphi$ and $\Phi$ are in general different. The difference can be demonstrated on contact of two metals with different Fermi energies. In equilibrium the electrochemical potentials of electrons in the two metals are equal, thus $\Phi$ is the same in both metals, whereas the Maxwell (or electrostatic) potentials $\varphi$ in the two metals are different, which compensates the difference in Fermi energies. Moreover, it is the electrostatic potential $\varphi$ (not $\Phi$) for which the Poisson equation holds.

\subsection{Electrochemical reactions}
Electrochemical reactions 
are driven by differences in electrochemical affinities, 
\begin{equation}
	\eA_r = \sum_\alpha \nu_{r\alpha}\mue_\alpha,
\end{equation}
where the electrochemical potentials $\mue_\alpha$ are evaluated near the electrode surfaces, see e.g. \cite{Dreyer-BV,guggenheim}. Equilibrium of an electrochemical reaction is then characterized by vanishing electrochemical affinity, 
\begin{equation}\label{eq.reactions.equilibrium}
	 \sum_\alpha \nu_{r\alpha}\mue_\alpha = 0.
\end{equation}

\subsection{Transport}
The flux of species $\alpha$ is typically proportional to the gradient of electrochemical potential of the species, 
\begin{equation}
\jj_\alpha = - \mathcal{D}_\alpha \nabla \mue_\alpha,
\end{equation}
see e.g. \cite{PKG,dGM}, where this relation is derived from microscopic theory.

Considering transport through a membrane, there is no net flux if the electrochemical potential of the transported species is the same on both sides of the membrane (neglecting possible coupling and crossover effects),
\begin{equation}\label{eq.membrane.equilibrium}
	\mue^A_\alpha = \mue^C_\alpha.
\end{equation}

Let us now discuss transport in battery electrolyte. According to Eq. \eqref{eq.mue}, the gradient of $\mue_\alpha$ can be split into the gradient of chemical potential $\mu_\alpha$, which depends on molalities of species in the mixture, and the gradient of Maxwell potential $\varphi$. Assuming perfect mixing (homogeneous concentrations), the condition of zero flux turns to a condition of no gradient of Maxwell electrostatic potential, 
\begin{equation}\label{eq.maxwell}
	\nabla \varphi = 0.
\end{equation}
The Maxwell electrostatic potential can be thus considered constant in the electrolyte (assuming no net flux and perfect mixing). Consequently, we assume that the electrochemical potential of species in the electrolyte is the same near the membrane and near the electrode, which greatly simplifies the modeling (since Poisson equation would have to be solved otherwise).

\subsection{Open-circuit voltage}
Open-circuit voltage (OCV) is an important quantity characterizing batteries and electrochemical cells, but it has no simple universal definition. Generally, it is the voltage measured by a potentiometer when no flux is passing through the cell. This definition can be ambiguous since the state of the cell is not specified. Perhaps the theoretically most appealing definition is that the electrochemical cell has been left undisturbed for sufficiently long time to reach thermodynamic equilibrium. Such equilibrium would be characterized by minimization of Gibbs free energy of all possible electrochemical reactions taking place in the cell, see \cite{Affro-saze}.

However, when measuring the OCV in practice, the cell is typically not let to relax to the complete equilibrium. The reason is two-fold, (i) it would take unrealistically long time and (ii) parasitic processes (like corrosion or crossover effects) would take place. The OCV is rather measured as the voltage once a plateau appears on the voltage vs. time plot. Therefore, it is necessary to consider the fastest processes affecting the practical OCV in the theoretical calculation, so that the measured plateau corresponds to equilibrium of these processes. 

OCV will then be defined as the difference in electric potential of electrons (proportional to the respective electrochemical potentials) at the positive and negative electrodes, 
\begin{equation}
	E = \Phi^P - \Phi^N.
\end{equation}
Using the equilibrium of electrochemical reactions \eqref{eq.reactions.equilibrium} and transport through the membrane \eqref{eq.membrane.equilibrium}, the OCV can be calculated as a function of activities of relevant species in the cell. 

The OCV dependence on activities of the species can be turned to a dependence on molalities or concentrations of the species once activity coefficients are specified. Usually, however, data for activity coefficients are lacking, which leads to simplifying assumptions of ideal mixtures (activity coefficients equal to unity). This simplification results in inaccuracies as noted in \cite{pavelka-jps}. 

To further express the OCV as a function of the state of charge\footnote{SOC is equal to 1 in a fully charged state while being equal to 0 in a completely discharged state.} (SOC), it is first necessary to calculate molalities (or concentrations) of the species in the cell as functions of the SOC. One thus has to choose a concrete reaction pathway and to calculate the relation between the charge passing through the membrane and changes of molalities of all relevant species in both electrolytes. 

Let us now demonstrate this thermodynamic derivation of the formula for OCV on several examples.

\section{Examples}
We shall commence with a simple $\rm{AgCl}$ cell connected with a standard hydrogen electrode. In this simple case both calculations of the Nernst relation coincide, and we discuss it for pedagogical clarity. Then we consider a zinc-air battery, where the thermodynamic approach overcomes the deficiencies of the formulas present in literature. Finally, the advantages of the thermodynamic approach are shown for vanadium redox-flow batteries, this example shows how the formula for OCV depends on the choice of the membrane (and the species transported through it).

\subsection{A simple example: The silver chloride electrode}
Let us first recall the simple case of an $\rm{AgCl}$ cell with a standard hydrogen electrode. The positive electrode material is $\rm Ag$, the main solute in the electrolyte is $\rm{HCl}$ and we assume that an ideal hydrogen electrode serves as a negative electrode. Moreover, we assume that a perfectly permselective cation-exchange membrane divides the positive and negative compartments, and that the electrolyte is well stirred in each compartment. The reaction taking place at the positive half-cell is 
\begin{subequations}\label{eq.AgCl.reactions}
\begin{equation}
	\rm{Ag + Cl^- \rightleftharpoons AgCl + e^-}
\end{equation}
while at the negative side we have
	\begin{equation}
		\rm{H^+ + e^- \rightleftharpoons \frac{1}{2}H_2}.
	\end{equation}
\end{subequations}
During discharge, electrons move from the anode (negative) through the external circuit to the cathode (positive), where they take part in the reduction reaction. Similarly, the hydrogen cations must be transferred from the anode, where they are formed, to the cathode through the electrolyte and membrane.

Assuming that the cell is in a steady state and that no current is passing through the external circuit (open-circuit condition), the electrochemical reactions do not proceed (have zero electrochemical affinities) and there is no ionic transport through the electrolyte or the membrane (zero gradient of electrochemical potential). 
Equilibrium in the positive electrode, $P$, reads
\begin{equation}
	0 = \mu_\mathrm{AgCl}^{P} + \mue_{e^-}^{P} - \mu_\mathrm{Ag}^{P} - \mue_\mathrm{Cl^-}^{P},
\end{equation}

This relation can be expanded using \eqref{eq.mue} and \eqref{eq.Phi}. The chemical potential of solids is typically near the standard value and, since the standard chemical potential of elements is zero, the chemical potential of silver is approximately zero, $\mu_\mathrm{Ag}^{P} \approx 0$ (see ~\ref{app.Gibbs}). Similarly, assuming the solution is saturated with $\rm AgCl$, we can deduce $\mu_\mathrm{AgCl}^{P} \approx \mu^\st_\mathrm{AgCl}$. The resulting positive half-cell potential is  

\begin{eqnarray}
	\phi^P &=& \frac{1}{F} \left( \mu_\mathrm{AgCl}^{P}-\mu_\mathrm{Ag}^{P} - \mue_\mathrm{Cl^-}^{P}\right)\nonumber\\
	&=& \frac{1}{F} \left(\mu^\st_\mathrm{AgCl}-\mu^\st_\mathrm{Cl^-}-RT\ln a^P_\mathrm{Cl^-} + F\varphi^{P}\right).
\end{eqnarray}

Electrochemical equilibrium in the negative electrode, $N$, is expressed by
\begin{subequations}
\begin{equation}
	0 = \frac{1}{2}\mu_\mathrm{H_2}^{N} - \mue_\mathrm{H^+}^{N} - \mue_{e^-}^{N},
\end{equation}
which can be rewritten as
\begin{equation}
	\phi^N = -\frac{1}{F}\left(\frac{1}{2}\mu_\mathrm{H_2}^{N} - \mue_\mathrm{H^+}^{N}\right) 
	= -\frac{1}{F}\left(RT\ln \sqrt{a^N_\mathrm{H^2}} - RT\ln a^N_\mathrm{H^+} - F\varphi^N\right),
\end{equation}
where split \eqref{eq.mue} was employed as well as the convention that the standard chemical potential of hydrogen ion is zero (in aqueous solution), $\mu^\st_\mathrm{H^+}=0$, as well as $\mu^\st_\mathrm{H_2}=0$.
\end{subequations}

Finally, assuming that $\rm H^+$ is the only species transported between the compartments (ideal cation-exchange membrane), equilibrium of transport through the membrane is expressed by
\begin{equation}\label{eq.Ag.H+}
	\mue_\mathrm{H^+}^{P} = \mue_\mathrm{H^+}^{N}.
\end{equation}

The voltage measured by a potentiometer is given by 
\begin{eqnarray}\label{eq.AgCl.OCV1}
	E = \Phi^{P} - \Phi^{N} = 
	\underbrace{\frac{1}{F} \left(\mu^\st_\mathrm{AgCl}-\mu^\st_\mathrm{Cl^-}\right)}_{=E^\st}-\frac{RT}{F}\ln\left(\frac{a^P_\mathrm{Cl^-}a^N_\mathrm{H^+}}{a^N_\mathrm{H^2}}\right)+\left(\varphi^P-\varphi^N\right).
\end{eqnarray}
The first term on the right hand side represents the standard potential and will be evaluated using thermodynamic data from \ref{app.Gibbs}. The second term represents the contribution of activities of the species. The third term, on the other hand, still has to be replaced by activities (and standard chemical potentials in the case of non-isothermal systems). To achieve that, the equilibrium of $\rm H^+$ transport through the membrane \eqref{eq.Ag.H+} can be rewritten, using expansion \eqref{eq.mue}, as

\begin{equation}\label{eq.H.eq}
\varphi^P-\varphi^N = \frac{RT}{F} \ln \left(\frac{a^N_\mathrm{H^+}}{a^P_\mathrm{H^+}}\right).
\end{equation}

Using this relation to simplify \eqref{eq.AgCl.OCV1} leads to the formula for OCV of the $\rm AgCl$ cell connected with a SHE
\begin{eqnarray}\label{eq.AgCl.OCV2}
	E = E^\st-\frac{RT}{F}\ln\left(\frac{a^P_\mathrm{Cl^-}a^P_\mathrm{H^+}}{a^N_\mathrm{H^2}}\right).
\end{eqnarray}

The standard cell potential then becomes $E^\st = \left(\mu^\st_\mathrm{AgCl}-\mu^\st_\mathrm{Cl^-}\right)/F = 0.223\ \mathrm{V}$.
Assuming dilute electrolytes, the ionic activities can be approximated by their molalities, 
\begin{equation}
	a^P_\mathrm{H^+} = \frac{b^P_\mathrm{H^+}}{b^\st} 
	\qquad\mbox{and}\qquad
	a^P_\mathrm{Cl^-} = \frac{b^P_\mathrm{Cl^-}}{b^\st}.
\end{equation}
Finally, the activity of hydrogen can be approximated by the ratio of its partial pressure to the standard pressure, $a_\mathrm{H_2}^N = p^N_\mathrm{H_2}/p^\st$.
The formula for OCV then becomes
\begin{equation}
	E = E^\st - \frac{RT}{F}\ln \frac{b_\mathrm{H^+}^P b_\mathrm{Cl^-}^P}{(b^\st)^2 \sqrt{\frac{p_\mathrm{H_2}^{N}}{p^\st}}},
\end{equation}
which is compatible with the usual formula form \cite{atkins}. We have included this example to demonstrate the thermodynamic derivation of open circuit voltage on the simple example where it coincides with the usual formula from literature.

\subsection{Zn-air redox flow battery}
Different equations for the OCV appear in the Zn-air battery literature \cite{zelger,schroder-ulrike,Stamm2017}. Although some authors do not explicitly present the OCV formula, they formulate the half-cell potentials and/or use them in the formulation of the Butler-Volmer equation. Here we show the correct thermodynamical derivation of the OCV formula for a zinc-air battery with an anion-exchange membrane and discuss some shortcomings of the equations found in the literature.

The most frequently used formula for OCV of Zn-air battery in literature, e.g. \cite{schroder-ulrike}, is
\begin{equation}\label{eq.OCV.usual}
 E_{usual} = E^P - E^N = E^\st - \frac{R T}{2F}\ln\frac{\sqrt{a^P_\mathrm{O_2}} a^P_\mathrm{H_2O} (c^N_\mathrm{OH^-})^4}{c^N_\mathrm{Zn(OH)^{2-}_4} (c^P_\mathrm{OH^-})^2},
\end{equation}
where the left hand side is the OCV while the right hand side consists of the standard voltage 
and a term dependent on activities of various species in the positive (P) and negative (N) electrolytes. We will show by careful application of thermodynamics that this relation should be corrected.

\begin{figure}[ht!]
\begin{center}
	\includegraphics[scale=0.3]{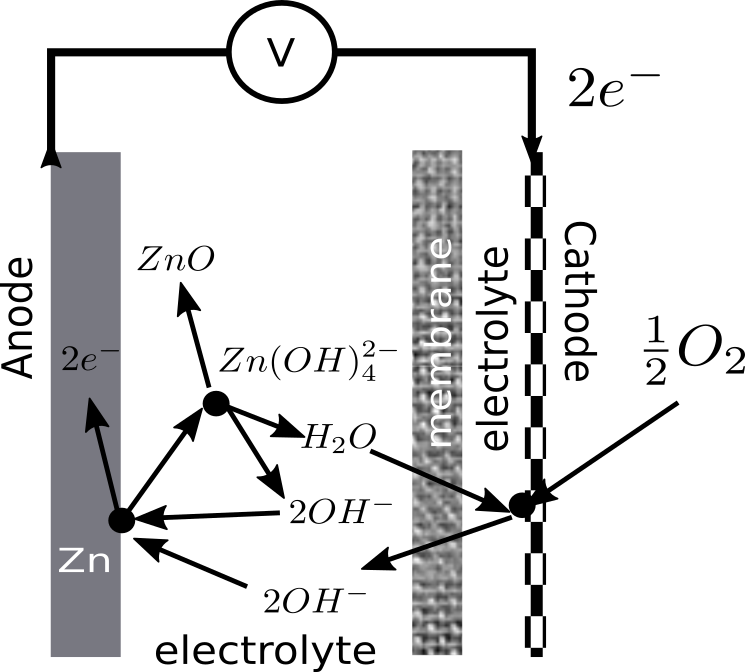}
	\caption{\label{fig.ZAB}Simplified scheme of a zinc-air battery.}
\end{center}
\end{figure}

\subsubsection{Electrochemical processes}
\begin{subequations}
The electrochemical reaction in the negative electrode of a zinc-air battery is
\begin{equation}\label{eq.chem.A}
\rm Zn + 4 OH^- \rightleftharpoons Zn(OH)^{2-}_4 + 2 e^-,
\end{equation}
and is accompanied by the precipitation of zinc oxide,
\begin{equation}\label{eq.Zn.ZnO}
\rm Zn(OH)^{2-}_4 \rightleftharpoons ZnO + 2OH^- + H_2O.
\end{equation}
On the positive half-cell, we have the following electrochemical reaction
\begin{equation}\label{eq.chem.C}
\rm \frac{1}{2} O_2 + H_2O + 2 e^- \rightleftharpoons 2 OH^-.
\end{equation}
Note that water and hydroxide ions are produced on one side and consumed on the other.
\end{subequations}
Assuming an ideal anion exchange membrane, $\rm{OH^-}$ is the only ion transported through the membrane. 


\subsubsection{Equilibrium conditions}
As in the previous example, we have to identify the equilibrium conditions at both electrodes and the membrane. Equilibrium of the negative electrode reaction is given by 
\begin{equation}\label{eq.Zn.A.Eq}
0 = \mue^N_\mathrm{Zn} + 4\mue^N_\mathrm{OH^-} - \mue^N_\mathrm{Zn(OH)^{2-}_4} - 2\mue^N_{e^-},
\end{equation}
which can be rewritten as
\begin{multline}\label{eq.Zn.Eq}
0 = \mu^\st_\mathrm{Zn} + 4\mu^\st_\mathrm{OH^-} + 4RT\ln a^N_\mathrm{OH^-} -4F \varphi^N \\
-\mu^\st_\mathrm{Zn(OH)^{2-}_4} -RT\ln a^N_\mathrm{Zn(OH)^{2-}_4} +2F\varphi^N  + 2F\Phi^N.
\end{multline}

Analogically, equilibrium of the positive electrode half-reaction reads
\begin{eqnarray}\label{eq.O2.Eq}
0 &=& \frac{1}{2} \mu^P_\mathrm{O_2} + \mu^C_\mathrm{H_2O} + 2 \mue^P_{e^-} - 2 \mue^P_\mathrm{OH^-}\nonumber\\
&=& \frac{1}{2} \mu^\st_\mathrm{O_2}+\frac{1}{2} RT \ln a^P_\mathrm{O_2} + \mu^\st_\mathrm{H_2O} + RT\ln a^P_\mathrm{H_2O} \nonumber\\
&&- 2 F\Phi^P 
- 2 \mu^\st_\mathrm{OH^-} - 2RT\ln a^P_\mathrm{OH^-} +2 F\varphi^P.
\end{eqnarray}

Equilibrium of the membrane transport is expressed by 
\begin{equation}\label{eq.Zn.OH}
	\mue^N_\mathrm{OH^-} = \mue^P_\mathrm{OH^-}.
\end{equation}

\subsubsection{Open-circuit voltage}
Combining \eqref{eq.Zn.Eq} and \eqref{eq.O2.Eq} the OCV of the battery is then given by
\begin{eqnarray}\label{eq.E.0}
E &=& \Phi^P - \Phi^N \\
&=&\frac{1}{2F}\left( \mu^\st_\mathrm{Zn} + 2\mu^\st_\mathrm{OH^-} - \mu^\st_\mathrm{Zn(OH)^{2-}_4} + \frac{1}{2}\mu^\st_\mathrm{O_2} + \mu^\st_\mathrm{H_2O}\right)\nonumber\\
&&+\frac{RT}{2F}\ln \left(\frac{\sqrt{a^P_\mathrm{O_2}} a^P_\mathrm{H_2O} (a^N_\mathrm{OH^-})^4}{(a^P_\mathrm{OH^-})^2 a^N_\mathrm{Zn(OH)^{2-}_4}}\right)\nonumber\\
&&+\varphi^P - \varphi^N.
\end{eqnarray}
The term involving the electrolyte potentials can be subtituted using the equilibrium of the transport of $\rm OH^-$ through the membrane \eqref{eq.Zn.OH} such that
\begin{equation}\label{eq.OH.eq}
\varphi^P-\varphi^N = \frac{RT}{2F} \ln \left(\frac{a^P_\mathrm{OH^-}}{a^N_\mathrm{OH^-}}\right)^2.
\end{equation}
Plugging this relation back into Eq. \eqref{eq.E.0} gives the final form of the general formula for OCV
\begin{equation}\label{eq.OCV}
E = E^\st  
+\frac{RT}{2F}\ln \left(\frac{\sqrt{a^P_\mathrm{O_2}} a^P_\mathrm{H_2O} (a^N_\mathrm{OH^-})^2}{ a^N_\mathrm{Zn(OH)^{2-}_4}}\right),
\end{equation}
where
\begin{equation}
E^\st =\frac{1}{2F}\left( \mu^\st_\mathrm{Zn} + 2\mu^\st_\mathrm{OH^-} - \mu^\st_\mathrm{Zn(OH)^{2-}_4} + \frac{1}{2}\mu^\st_\mathrm{O_2} + \mu^\st_\mathrm{H_2O}\right)=1.59,
\end{equation}
using the standard chemical potential values compiled in \ref{app.Gibbs}. 

\subsubsection{Comparison with the usual OCV}
Formula \eqref{eq.OCV} differs from the usual formula \eqref{eq.OCV.usual} by those terms coming from the equilibrium of transport of $\mathrm{OH^-}$ through the membrane \eqref{eq.OH.eq}. These terms can be referred to as the membrane (or Donnan) potential. By including them, the formula \eqref{eq.OCV} becomes more precise than the usual formula. It is important to go back to the thermodynamic roots of Nernst relation in the case of zinc-air batteries.

However, note that, when using a porous separator in the battery instead of the ion-exchange membrane, practically any species small enough to pass through the pores can be transported between anode and cathode sides. To obtain a relation for $\varphi^P-\varphi^N$ one has to determine the electrochemical process that equilibrates first across the separator. For instance, another charged species can be first to equilibrate, and Eq. \eqref{eq.Zn.OH} has to be then replaced by the analogical equilibrium equation for the species. In some situations, one can also assume that the electrolyte is perfectly mixed in the whole system which using \eqref{eq.maxwell} leads to 
\begin{equation}\label{eq.separator}
	\varphi^P-\varphi^N = 0
\end{equation}

Even though most works in Zn-air systems use separators instead of membranes, several of the formulas present in the literature \cite{zelger,schroder-ulrike,Stamm2017} have small deficiencies that would be prevented using the thermodynamic derivation described here. For example, \cite{zelger} presents a formula where the activity term has an opposite sign. In \cite{schroder-ulrike}, the concentrations of $\rm O_2$ and $\rm H_2O$ are used instead of their activities. For pure liquids (or in this case solvents) and dissolved gases the relation between activities and concentrations required a different approach to the reference concentration. The proper approach for the dissolved gas concentration was identified by \cite{Stamm2017}, where they properly treat the $\rm O_2$ concentration by using as reference state the concentration in equilibrium with $\rm O_2$ gas in standard conditions given by Henry's law.

\subsubsection{Including corrosion}
The reaction scheme \eqref{eq.chem.A} and \eqref{eq.chem.C} does not take into side reactions happening on the electrodes. In reality, at the equilibrium potential of the zinc electrode the hydrogen evolution reaction (HER) will also take place
\begin{equation}\label{eq.chem.H2O}
	\rm 2 H_2O + 2 e^- \rightleftharpoons H_2 + 2OH^-.
\end{equation}
This is a corrosion process that consumes active material (zinc) and shifts the OCV. The half-cell potential in this situation is usually referred as mixed potential.

The fast equilibrium of the corrosion pair \eqref{eq.chem.A} and \eqref{eq.chem.H2O} is characterized by equality of their currents (no net current). Reaction rates are typically expressed by the Butler--Volmer equation
\begin{equation}\label{eq.BV}
	j = j_0 \left(e^{\alpha\eA/RT}-e^{-(1-\alpha) \eA/RT}\right),
\end{equation}
see e.g. \cite{pavelka-jps} for a thermodynamic origin of this equation or \cite{atkins} for a kinetic origin. 
The charge transfer coefficient $\alpha$ describes whether the transition state is closer to the oxidized or reduced species, typically it is assumed to be equal to $1/2$, \cite{atkins}. The exchange current $j_0$ prefactor can be either constant or dependent on concentrations of the reactants and products, c.f. \cite{newman} or \cite{PKG} for the Boltzmann equation, and it is positive in the direction of oxidation. 

In general, the mixed potential can be determined by numerical methods using the equality of currents for zinc dissolution and hydrogen oxidation. %
 Nonetheless, in some cases Butler-Volmer equation can be simplified thus leading to analytical solutions of the mixed potential. Such example, is discussed in \ref{app.corr}.

\subsection{All vanadium flow cell} 
The vanadium redox flow battery (VRFB) was analyzed in \cite{pavelka-jps} and \cite{Catalina}. Let us briefly recall the calculation in a simplified form. We shall consider two cases: VRFB with a cation-exchange (catex) membrane and VRFB with an anion-exchange (anex) membrane. The OCV formula will be different in the two cases.

The main electrochemical reactions taking place in VRFB can be summarized as
\begin{subequations}\label{eq.VRFB.reactions}
	\begin{equation}
		\rm VO_2^+ + e^- + 2H^+ \rightleftharpoons VO^{2+} + H_2O
	\end{equation}
	on the positive side
	and
	\begin{equation}
		\rm V^{2+} \rightleftharpoons V^{3+}+e^-
	\end{equation}
	on the negative side.
\end{subequations}
Furthermore, we should also consider ionic transport through the membrane, $\rm H^+$ in the catex case and $\rm HSO_4^-$ in the anex case. Let us now discuss these two cases separately.

\subsubsection{Cation-exchange membrane}
The equations expressing equilibrium of the electrochemical reactions are
\begin{subequations}\label{eq.VRFB.reactions.eq}
	\begin{equation}
		\mue^P_\mathrm{VO_2^+} -F \Phi^P + 2\mue^P_\mathrm{H^+} = \mue^P_\mathrm{VO^{2+}} + \mu^P_\mathrm{H_2O}
	\end{equation}
	and
	\begin{equation}
		\mue^N_\mathrm{V^{2+}} = \mue^N_\mathrm{V^{3+}} - F\Phi^N.
	\end{equation}
\end{subequations}
Equilibrium of the ionic transport through the membrane means that
\begin{equation}
	\mue^N_\mathrm{H^+} = \mue^P_\mathrm{H^+},
\end{equation}
which can be rewritten as
\begin{equation}\label{eq.VRFB.membrane.catex}
	\varphi^P - \varphi^N = \frac{RT}{F}\ln \frac{a^N_\mathrm{H^+}}{a^P_\mathrm{H^+}},
\end{equation}
where $\varphi^{P,N}$ are the Maxwell potentials in the positive and negative electrolytes, respectively.

The OCV can be then expressed as
\begin{eqnarray}\label{eq.VRFB.OCV}
	E &=& \Phi^P-\Phi^N = 
		\frac{1}{F}\left(\mue^P_\mathrm{VO_2^+} + 2\mue^P_\mathrm{H^+} - \mue^C_\mathrm{VO^{2+}} - \mu^P_\mathrm{H_2O}\right)
		+\frac{1}{F}\left(\mue^N_\mathrm{V^{2+}} - \mue^N_\mathrm{V^{3+}}\right)\nonumber\\
	&=& \frac{1}{F}\left(\mu^\st_\mathrm{VO_2^+} - \mu^\st_\mathrm{VO^{2+}} - \mu^\st_\mathrm{H_2O} +\mu^\st_\mathrm{V^{2+}} - \mu^\st_\mathrm{V^{3+}}\right)\nonumber\\
	&&+\frac{RT}{F} \ln \frac{a^P_\mathrm{VO_2^+} (a^P_\mathrm{H^+})^2 a^N_\mathrm{V^{2+}}}{a^P_\mathrm{VO^{2+}} a^P_\mathrm{H_2O} a^N_\mathrm{V^{3+}}} \nonumber\\
	&&+\varphi^P-\varphi^N,
\end{eqnarray}
where the split \eqref{eq.mue} was used. Note the last line, where the difference in Maxwell electrostatic potentials remains. This is caused by the form of Eqs. \eqref{eq.VRFB.reactions}, where the charge on the left hand sides is equal to the charge on the right hand sides, but is not zero. 
To get rid of the Maxwell potential difference, we have to use the membrane equilibrium equations \eqref{eq.VRFB.membrane.catex}, which leads to 
\begin{equation}\label{eq.VRFB.OCV.catex}
	E = E^\st
	+\frac{RT}{F} \ln \frac{a^P_\mathrm{VO_2^+}  a^N_\mathrm{V^{2+}} a^P_\mathrm{H^+} a^N_\mathrm{H^+}}{a^P_\mathrm{VO^{2+}} a^C_\mathrm{H_2O} a^N_\mathrm{V^{3+}}},
\end{equation}
where the standard cell potential is
\begin{equation}
	E^\st =  \frac{1}{F}\left(\mu^\st_\mathrm{VO_2^+} - \mu^\st_\mathrm{VO^{2+}} - \mu^\st_\mathrm{H_2O} +\mu^\st_\mathrm{V^{2+}} - \mu^\st_\mathrm{V^{3+}}\right) = 1.256\ \mathrm{V}, 
\end{equation}
taking values of the standard Gibbs energies of formation from \cite{Wagman}. Formula \eqref{eq.VRFB.OCV.catex} is the formula for the OCV in the catex case. To compare it with experimental data, which are usually expressed in terms of SOC, the activities have to be expressed in terms of molalities.

\subsubsection{Anion-exchange membrane}
\begin{figure}
	\begin{center}
		\includegraphics[scale=0.8]{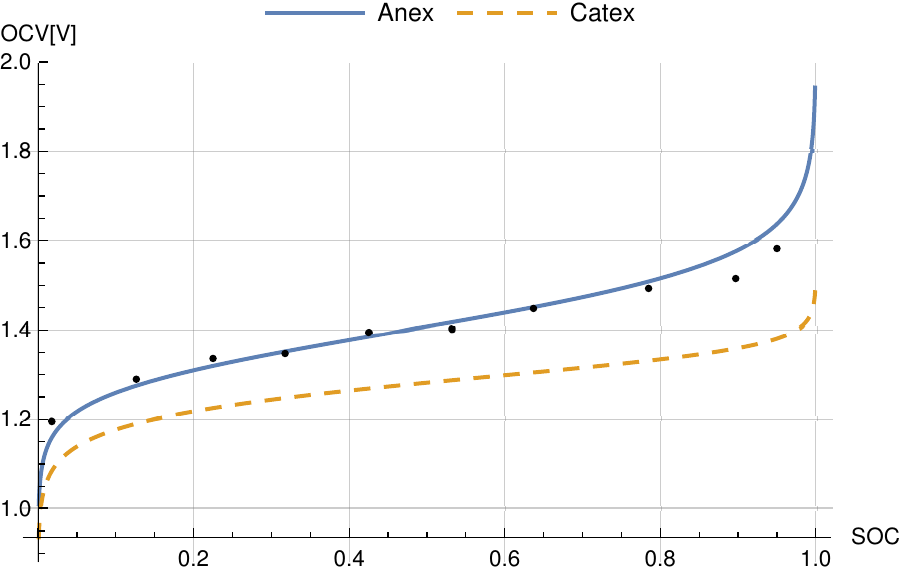}
		\caption{\label{fig.VRFB}Comparison of the catex formula \eqref{eq.VRFB.OCV.catex} and anex formula \eqref{eq.VRFB.OCV.anex} with experimental data for an anex membrane (data re-used from \cite{pavelka-jps}). The catex formula, which is close to the usual Nernst relation for VRFBs is clearly deficient. For comparison of the formula for catex membranes with experimental data see \cite{pavelka-jps}.}
	\end{center}
\end{figure}

We shall now discuss VRFBs with anion exchange membranes through which $\rm HSO_4^-$ ions are transported. The equilibrium equations \eqref{eq.VRFB.reactions.eq} are the same as in the anex case. The difference is in the ion transport as Eq. \eqref{eq.VRFB.membrane.catex} changes to
\begin{equation}
	\mue^N_\mathrm{HSO_4^-} = \mue^P_\mathrm{HSO_4^-},
\end{equation}
which can be rewritten as
\begin{equation}\label{eq.VRFB.membrane.anex}
	\varphi^P - \varphi^N = \frac{RT}{F}\ln \frac{a^P_\mathrm{HSO_4^-}}{a^N_\mathrm{HSO_4^-}}.
\end{equation}
The OCV \eqref{eq.VRFB.OCV} then turns to
\begin{equation}\label{eq.VRFB.OCV.anex}
	E = E^\st
	+\frac{RT}{F} \ln \frac{a^P_\mathrm{VO_2^+}  a^P_\mathrm{V^{2+}} (a^P_\mathrm{H^+})^2 a^C_\mathrm{HSO_4^-}}{a^P_\mathrm{VO^{2+}} a^P_\mathrm{H_2O} a^N_\mathrm{V^{3+}} a^N_\mathrm{HSO_4^-} }.
\end{equation}
Again, to compare it with the experimental dependence of OCV on SOC accurately, one needs to express the activities, exploiting e.g. \cite{VRFB-activities}, in terms of molalities. Neglecting the activity coefficients, the comparison is shown in Fig. \ref{fig.VRFB}, where data were taken from \cite{pavelka-jps}.

Note also that the formula for OCV is different in the anex case than in the catex case, such a difference can not be seen in the naive construction of Nernst relation.

The thermodynamic formula for OCV was found different from the usual one available in the literature \cite{knehrkumbur}. Moreover, it was found that the formula for OCV depends on the choice of the membrane, more precisely on the species that is transported through the membrane \cite{pavelka-jps}.  Although the electrochemical equations are the same, the equilibrium equation for transport through the membrane is different, as well as the resulting Nernst relation.

\section{Conclusions}
The simplified way towards Nernst relation is extensively used in the literature. Nonetheless, that way can lead to inaccurate conclusions when analyzing and modelling a battery system for instance in zinc-air and vanadium redox flow batteries. The thermodynamic back-to-the-roots derivation of the Nernst relation provides a more robust way towards the open-circuit voltage, which takes into account all the relevant interphases.

In particular, in systems with ion-exchange membranes the OCV depends on the species transported through the membrane. In vanadium redox flow batteries, where both anion-exchange and cation-exchange membranes can be used, two different Nernst relations are obtained.

\section*{Acknowledgments}
This project has received funding from the European Union’s Horizon 2020 research and innovation programme under the Marie Sk\l{}odowska-Curie Grant Agreement no. 765289.
MP was supported by Czech Science Foundation, project no.  17-15498Y, and by Charles University Research program No. UNCE/SCI/023.

\section*{References}

\appendix
\section{Gibbs energies of formation}\label{app.Gibbs}

The standard chemical potentials can be taken as equal to the standard Gibbs energies of formation $\Delta_f G^\st$, \cite{atkins,Wagman}. The standard Gibbs energy of formation of elements is zero by definition, as well as $\Delta_f G^\st$ of $\rm{H^+}$ in water. The following table shows the standard Gibbs energy of formation of the chemical compounds used in this paper following \cite{Wagman} (except where indicated otherwise). The reference state corresponds to a temperature of $298.15\ \mathrm{K}$, a pressure of $1\ \mathrm{bar}$, and a molality of $1\ \mathrm{mol\ kg^{-1}}$.

\begin{table}[ht!]
\centering
\begin{tabular}{| c | c |}
\hline
Compound & $\Delta G_f^\st\ \mathrm{[kJ\ mol^{-1}]}$ \\
\hline
$\rm H^+$ & $0$  \\ 
$\rm Cl^-$ & $-131.288$  \\
$\rm AgCl$ & $-109.789$  \\
$\rm OH^-$ & $-157.244$   \\
$\rm H_2O$ & $-237.129$   \\
$\rm Zn(OH)_4^{2-}$ & $-858.52$ \\
$\rm V^{2+}$ & $-217.6$ \\
$\rm V^{3+}$ & $-242.3$ \\
$\rm VO^{2+}$ & $-446.4$ \cite{hill1970} \\
$\rm VO_2^{+}$ & $-587.0$ \cite{hill1970} \\
$\rm HBr$ & $-103.96$ \\
\hline
\end{tabular}
\end{table}


\section{Corrosion in zinc electrodes}\label{app.corr}

Under certain conditions, an analytical solution for the mixed potential can be obtained. In the case of high $\eA$, the Butler--Volmer equation \eqref{eq.BV} can be simplified to the Tafel equation
\begin{equation}
	j = j_0 e^{(1-\alpha)\eA/RT}
\end{equation}
while for the low $\eA$ case it turns to a linear relation
\begin{equation}
	j = j_0 \frac{\eA}{RT}
\end{equation}
between the current and the overpotential. Note that the charge transfer coefficient $\alpha$ disappears in the low-current limit.

Zinc electrodes are reported to have fast kinetics and relatively high hydrogen overpotentials \cite{Cachet1982,Chen1991,Dundalek2017}. Consequently, the HER is usually in the Tafel regime for zinc systems (using the typical value $\alpha=1/2$) while the zinc dissolution is in the linear regime, that is
\begin{subequations}
	\begin{eqnarray}
		j^\mathrm{HER} &=& -j^\mathrm{HER}_0 e^{\eA_\mathrm{HER}/(2RT)}\\
		j^\mathrm{Zn} &=& j^\mathrm{Zn}_0 \eA_\mathrm{Zn}/(RT)
	\end{eqnarray}
\end{subequations}
for reactions \eqref{eq.chem.H2O} and \eqref{eq.chem.A}, respectively. The electrochemical affinities are defined as
\begin{subequations}
	\begin{eqnarray}
		\eA_\mathrm{HER} &=& 2\mu_\mathrm{H_2O} - 2F\Phi^A -\mu_\mathrm{H_2} - 2\mue^A_\mathrm{OH^-}\\
		\eA_\mathrm{Zn} &=& \mu^A_\mathrm{Zn} +4\mue^A_\mathrm{OH^-} -\mue^A_\mathrm{Zn(OH)^{2-}_4} +2F\Phi^A.
	\end{eqnarray}
\end{subequations}
The mixed potential equilibrium corresponds, assuming the same electroactive area for both reactions, to an equality of currents,
\begin{equation}
	j^\mathrm{corr}= j^\mathrm{Zn}=-j^\mathrm{HER},
\end{equation}
which can be rewritten as a nonlinear algebraic equation for $\Phi^A-\varphi^A$,
\begin{multline}
\frac{1}{RT}\left(\mu^N_\mathrm{Zn}+4\mu^N_\mathrm{OH^-}-\mu^N_\mathrm{Zn(OH)_4^{2-}}+2F (\Phi^N-\varphi^N)\right)	\\
	= \frac{j_0^\mathrm{HER}}{j_0^\mathrm{Zn}} e^{\left(\frac{2\mu^N_\mathrm{H_2O} -\mu^N_\mathrm{H_2}-2\mu^N_\mathrm{OH^-}}{2RT}\right)}
	e^{-\frac{F}{RT}(\Phi^N-\varphi^N)}.
\end{multline}

Introducing dimensionless quantities $\hat{A}_\mathrm{Zn} = \frac{\mu^N_\mathrm{Zn}+4\mu^N_\mathrm{OH^-}-\mu^N_\mathrm{Zn(OH)_4^{2-}}}{RT}$, 
$\hat{A}_\mathrm{HER} = \frac{2\mu^N_\mathrm{H_2O} -\mu^N_\mathrm{H_2}-2\mu^N_\mathrm{OH^-}}{2RT}$, $\hat{j_0}=\frac{j_0^\mathrm{HER}}{j_0^\mathrm{Zn}}$ and $\hat{\Phi} = (\Phi^N-\varphi^N)\frac{F}{RT}$, results in the {non-dimensional} equation
\begin{equation}\label{eq.Ban.non}
	\hat{\Phi} 	
	= -\frac{1}{2}\hat{A}_\mathrm{Zn}
	+j_0 e^{\hat{A}_\mathrm{HER}}
	e^{-\hat{\Phi}}.
\end{equation}
This non-dimensional nonlinear algebraic equation can be solved by the Banach fixed point iteration (proof of convergence can be found in \ref{sec.Banach}), using $\hat{\Phi}^{(n)} = T(\hat{\Phi}^{(n-1)})$ with mapping $T(\hat{\Phi})$ defined as the right hand side of Eq. \eqref{eq.Ban.non}. Taking the zero-th approximation $\hat{\Phi}^{0} = \hat{A}_\mathrm{Zn}$, which corresponds to the standard hydrogen electrode, the first approximation reads
\begin{subequations}
	\begin{equation}
		\hat{\Phi}^{(1)} = 
	 -\frac{1}{2}\hat{A}_{Zn}
	+j_0 e^{\hat{A}_\mathrm{HER}}
		e^{-\hat{A}_\mathrm{Zn}}.
	\end{equation}
	while the following
	\begin{equation}
		\hat{\Phi}^{(i)} = 
	 -\frac{1}{2}\hat{A}_\mathrm{Zn}
	+j_0 e^{\hat{A}_\mathrm{HER}}
		e^{-\hat{\Phi}^{(i-1)}}.
	\end{equation}
\end{subequations}
Let us denote the final solution as $\hat{\Phi}^N$.

The formula for OCV is defined as the difference of electrochemical potentials of electrons as before,
\begin{eqnarray}
	E &=& \Phi^P - \Phi^N = \frac{1}{2F}\left(\frac{1}{2}\mu^C_\mathrm{O_2} + \mu^C_\mathrm{H_2O} - 2\mue^C_\mathrm{OH^-}\right)\nonumber\\
	&&- \frac{RT}{F}\hat{\Phi}^N -\varphi^N\nonumber\\
	 &=& \frac{1}{2F}\left(\frac{1}{2}\mu^C_\mathrm{O_2} + \mu^C_\mathrm{H_2O} - 2\mu^C_\mathrm{OH^-} - 2 RT \hat{\Phi}^N\right)\nonumber\\
	&&+\varphi^P-\varphi^N,
\end{eqnarray}
where $\hat{\Phi}$ is the result of the fixed point iteration. Note that the last expression consists of a chemical contribution and electrical contribution, which is to be replaced by a transport equilibrium condition, e.g. formula \eqref{eq.Zn.OH}.

Let us assume a low hydrogen pressure in the system, and a solution of $6\ \mathrm{M\ KOH}$ with unit activities for rest of the species. The negative half-cell potential $\phi^N$ resulting from the fixed-point iteration for different values of $\hat{j_0}$ is shown in Table~\ref{tab.corr}.

\begin{table}[ht!]
\centering
\caption{\label{tab.corr}Values.}
\begin{tabular}{| c | c | c|}
\hline
\rule{0pt}{15pt}
$\hat{j_0}$ & $\phi^N$ & Nr. iterations\\
\hline
$10^{-3}$ & $-1.261$ & 12 \\
$10^{-4}$ & $-1.278$ & 2 \\
$10^{-5}$ & $-1.281$  & 1\\
$10^{-6}$ & $-1.282$  & 0\\
\hline
\end{tabular}
\end{table}

%
%
%
%
%
%
%
%

\section{Banach fixed point theorem} \label{sec.Banach}
Having a Banach space $X$ (e.g. real numbers with a metric given the absolute value) and a contraction mapping $T:X\rightarrow X$ satisfying
\begin{equation}
	\exists q\in(0,1)\quad \mbox{for which}\quad |T(x)-T(y)|\leq q \cdot |x-y|\quad \forall x,y\in X,
\end{equation}
there is only one fixed point of the mapping, $T(\bar{x}) = \bar{x}$, and it is the limit of iterations $x^{(n)} = T(x^{(n-1)})$. This is the Banach contraction theorem.

In order to apply it to Eq. \eqref{eq.Ban.non}, we have to verify that the mapping 
\begin{equation}
	T(\hat{\Phi}) = -\frac{1}{2}\hat{A}_{Zn}
	+j_0 e^{\hat{A}_{HER}}
	e^{-\hat{\Phi}}
\end{equation}
is really a contraction. Therefore, we calculate (assuming $\hat{\Phi}_1 \geq \hat{\Phi}_2$ without any loss of generality)
\begin{eqnarray}
	|T(\hat{\Phi}_1)-T(\hat{\Phi})_2| &= &
	j_0 e^{\hat{A}_{HER}} 
	\left|e^{-\hat{\Phi}_1}-e^{-\hat{\Phi}_2}\right|\\
	&=&
	j_0 e^{\hat{A}_{HER}} 
	\left|\int_{\hat{\Phi}_2}^{\hat{\Phi}_1} -e^{-\hat{\Phi}}d\hat{\Phi}\right|\nonumber\\
	&=&
	j_0 e^{\hat{A}_{HER}} 
	\int_{\hat{\Phi}_2}^{\hat{\Phi}_1} \nonumber\\
	&\leq&
	j_0 e^{\hat{A}_{HER}} 
	\int_{\hat{\Phi}_0-\epsilon}^{\hat{\Phi}_0+\epsilon} e^{-\hat{\Phi}} d\hat{\Phi} = 
	j_0 e^{\hat{A}_{HER}} \cdot 2 e^{-\hat{\Phi}_0} \sinh(\epsilon)\nonumber
\end{eqnarray}
where it was assumed that $\hat{\Phi}_0$ is in the vicinity of a value $\hat{\Phi}_0$, $\hat{\Phi}\in (\hat{\Phi}_0-\epsilon, \hat{\Phi}_0+\epsilon)$. If $j_0 e^{\hat{A}_\mathrm{HER}}$ is low enough, then the mapping is indeed a contraction, i.e. $|T(\hat{\Phi}_1)-T(\hat{\Phi})_2|\leq q$ for some $q<1$. This is usually satisfied because of the small value of $j_0$ given by the small value of $j_0^\mathrm{HER}$, and the iteration then converges.

\section{Hydrogen-Bromine redox flow battery}\label{app.H2Br2}
Let us now discuss an idealized hydrogen-bromine battery, see e.g. \cite{Tucker-Cho-Weber}.

\subsection{Electrochemical processes}
The main electrochemical reactions taking place in hydrogen-bromine redox batteries are
\begin{subequations}
	\begin{equation}
		\rm Br_2(aq) + 2e^- \rightleftharpoons 2Br^-(aq)
	\end{equation}
	and
	\begin{equation}
		\rm 2Br^-(aq) + 2H^+ \rightleftharpoons 2HBr
	\end{equation}
	on the cathode side and
	\begin{equation}
		\rm H_2(g) \rightleftharpoons 2H^+ + 2e^-
	\end{equation}
	on the anode side. 
\end{subequations}

A cation exchange membrane then facilitates transport of $\rm H^+$ from the anode to the cathode.

\subsection{Equilibrium conditions}
Equilibrium of the electrochemical reactions is expressed as
\begin{subequations}
	\begin{equation}
		\mu^P_\mathrm{Br_2} - 2F \Phi^P = 2\mue^P_\mathrm{Br^-},
	\end{equation}
	\begin{equation}
		\mue^P_\mathrm{Br^-} + \mue^P_\mathrm{H^+} = \mu^P_\mathrm{HBr}
	\end{equation}
	and
	\begin{equation}
		\mu^N_\mathrm{H_2} = 2\mue^N_\mathrm{H^+} - 2\Phi^N.
	\end{equation}
\end{subequations}
Equilibrium of the membrane transport reads
\begin{equation}
	\mue^N_\mathrm{H^+} = \mue^P_\mathrm{H^+}.
\end{equation}

\subsection{Open circuit voltage}
The open circuit voltage is then calculated as 
\begin{eqnarray}\label{eq.OCV.HBr}
	E &=& \Phi^P-\Phi^N = \frac{1}{2F}\left(\mu^P_\mathrm{Br_2} - 2(\mu^P_\mathrm{HBr}-\mue^P_\mathrm{H^+})+\mu^N_\mathrm{H_2} - 2\mue^N_\mathrm{H^+}\right)\nonumber\\
	&=& \underbrace{\frac{1}{2F}\left(\mu^\st_\mathrm{Br_2} - 2 \mu^\st_\mathrm{HBr} + \mu^\st_\mathrm{H_2}\right)}_{=E^\st}
	+\frac{RT}{2F} \ln \frac{a^N_\mathrm{H_2} a^P_\mathrm{Br_2}}{(a^P_\mathrm{HBr})^2}.
\end{eqnarray}
Since the chemical potential of elements at standard conditions is zero, the standard cell potential becomes	$E^\st = -\mu^\st_\mathrm{HBr}/{F} = 1.077\ \mathrm{V}$. Using the dissociation equilibrium of $\rm HBr$, 
\begin{equation}
	\mu^P_\mathrm{HBr} = \mue^P_\mathrm{H^+}  + \mue^P_\mathrm{Br^-}, 
\end{equation}
from which it follows that $\mu^\st_\mathrm{HBr}  = \mu^\st_\mathrm{Br^-}$ and $a^C_\mathrm{HBr} = a^P_\mathrm{H^+} \cdot a^P_\mathrm{Br^-}$, formula \eqref{eq.OCV.HBr} can be equivalently rewritten as
\begin{equation}
	E 
	= E^\st
	+\frac{RT}{2F} \ln \frac{a^N_\mathrm{H_2} a^P_\mathrm{Br_2}}{(a^N_\mathrm{H^+})^2  (a^P_\mathrm{Br^-})^2},
\end{equation}
which is the same as the usual Nernst relation, e.g. \cite{Kreutzer}. 
The usual OCV is compatible with thermodynamics in the case of hydrogen-bromine batteries.

\end{document}